\documentclass[a4paper]{jpconf}
\usepackage{graphicx}
\usepackage{dcolumn}
\usepackage{bm}
\usepackage{color}
\usepackage{amsmath}
\usepackage{amssymb}
\begin{document}
\title{Spin-Triplet Superconductivity Induced by Longitudinal Ferromagnetic Fluctuations in UCoGe: Theoretical Aspect  }

\author{Y. Tada,$^1*$ S. Fujimoto,$^2$, N. Kawakami,$^2$ T. Hattori,$^2$ Y. Ihara,$^3$ K. Ishida,$^2$ K. Deguchi,$^4$ N. K. Sato,$^4$ and I. Satoh$^5$}

\address{$^1$ Institute for Solid State Physics, the University of Tokyo, Kashiwa, Chiba 277-8581, Japan\\
$^2$Department of Physics, Graduate School of Science, Kyoto University, Kyoto 606-8502, Japan\\
$^3$Department of Physics, Graduate School of Science, Hokkaido University,Sapporo 060-0810, Japan\\
$^4$Department of Physics, Graduate School of Science, Nagoya University, Nagoya 464-8602, Japan\\
$^5$Institute for Materials Research, Tohoku University, Sendai 980-8577 Japan\\}

\ead{tada@issp.u-tokyo.ac.jp}

\newcommand{\vecc}[1]{\mbox{\boldmath $#1$}}
\newcommand{\red}{\textcolor{red}}
\newcommand{\blue}{\textcolor{red}}

\begin{abstract}
Identification of pairing mechanisms leading to the unconventional 
superconductivity realized in copper-oxide, heavy-fermions, 
and organic compounds is one of the most challenging 
issues in condensed-matter physics.
Clear evidence for an electron-phonon mechanism in conventional 
superconductors is seen by the isotope effect on 
the superconducting transition temperatures $T_{\rm SC}$, 
since isotopic substitution varies the phonon frequency 
without affecting the electronic states. 
In unconventional superconductors, magnetic fluctuations 
have been proposed to mediate superconductivity, 
and considerable efforts have been made to unravel relationships 
between normal-state magnetic fluctuations 
and superconductivity.
Here, we show that characteristic experimental results on the 
ferromagnetic (FM) superconductor UCoGe 
($T_{\rm Curie} \sim 2.5 $ K and $T_{\rm SC} \sim 0.6$ K) 
can be understood consistently within a scenario of 
the spin-triplet superconductivity 
induced by FM spin fluctuations. 
Temperature and angle dependencies of the upper critical magnetic field 
of the superconductivity ($H_{c2}$) 
are calculated on the basis of the above scenario by solving the Eliashberg equation.
Calculated $H_{c2}$ well agrees with the characteristic experimental 
results observed in UCoGe.
This is a first example that FM fluctuations are shown to 
be a pairing glue of superconductivity. 
\end{abstract}

\section{Introduction}
\label{sec:intro}
The discovery of superconductivity in ferromagnetic (FM) UGe$_2$
 under pressure gave great impact for condensed-matter community~\cite{pap:Saxena2000}, 
since ferromagnetism and superconductivity have been 
considered to be mutually exclusive phenomena.
After the discovery, 
superconductivity has been found in several U-based FM 
compounds~\cite{pap:Aoki2001,pap:Huy2007}.    
UCoGe is one of the compounds among the FM superconductors discovered so far 
which can be readily explored in experiments, 
because of its high superconducting (SC) transition 
temperature ($T_{\rm SC}$) and 
low Curie temperature ($T_{\rm Curie}$) at 
ambient pressure.
Actually, important and intriguing experimental 
results have been reported for UCoGe~\cite{pap:Gasparini_review,pap:Aoki_review}, and some of them are summarized as follows. 
\begin{enumerate}
\item
Microscopic measurements of $\mu$SR [Fig.1(a)]~\cite{pap:deVisser2009}  and Co-NQR [Fig.1(b)]~\cite{pap:Ohta2010} have shown that 
the superconductivity occurs within the FM region, 
resulting in homogeneous microscopic coexistence of 
the ferromagnetism and superconductivity.
The superconductivity is also found in the paramagnetic state and
$T_{\rm SC}$ shows a broad maximum around a critical pressure of ferromagnetism [Fig.1(c)]~\cite{pap:Huy2007}, which is in sharp contrast with UGe$_2$ where the SC is seen only in the
FM phase~\cite{pap:Saxena2000}.
\begin{figure}[h]
\begin{center}
\includegraphics[width=30pc]{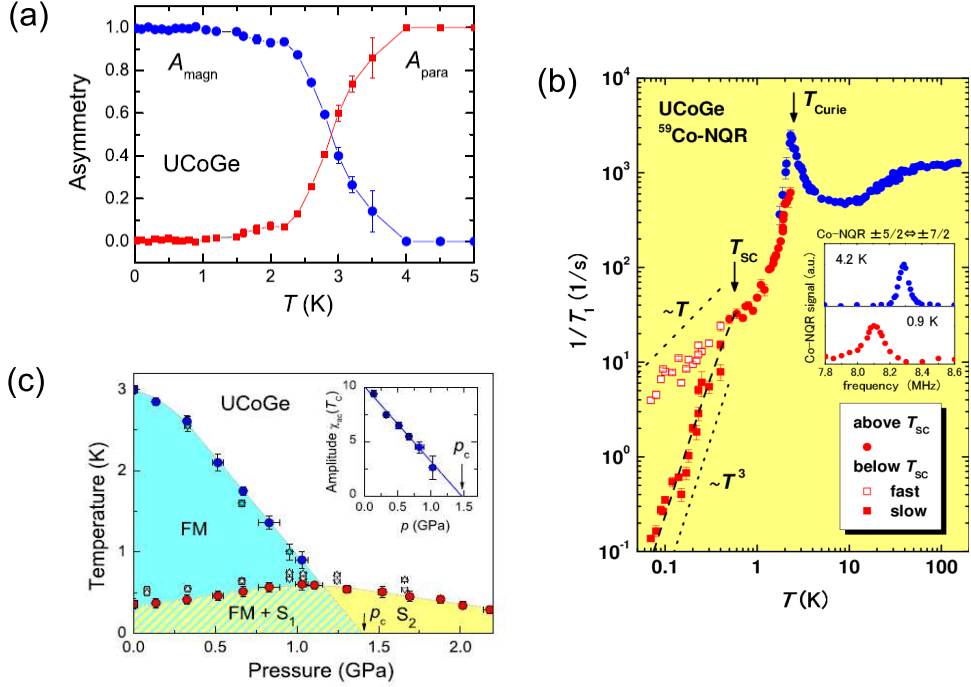}
\end{center}
\caption{\label{fig:1}(a)Temperature variation of the normalized asymmtries ($A_{\rm magn}+A_{\rm para} =1$) of the FM and paramagnetic phases measured with zero-field muon-spin rotation on UCoGe \cite{pap:deVisser2009}. Below $\sim 1.5$ K, magnetism is observed in the whole sample volume although superconductivity occurs at $\sim$ 0.8 K, indicative of coexistence of ferromagnetism and superconductivity on a microscopic scale. 
(b) Temperature dependence of  $^{57}$Co $1/T_1$ in the single-crystal sample \cite{pap:Ohta2010}. $1/T_1$ above 2.3 K, shown by the blue dots, was measured at the paramagnetic (PM)-frequency (8.3 MHz) peak arising from the $\pm 5/2 \leftrightarrow \pm 7/2$ transitions, and  $1/T_1$ below 2.3 K, shown by the red dots, was measured at the FM-frequency (8.1 MHz) peak shifted from the PM peak due to the appearance of the internal field at the Co site. The inset shows the NQR signal arising from the $\pm 5/2 \leftrightarrow \pm 7/2$ transitions obtained at 4.2 K (PM) and 0.9 K (FM). 
Two $1/T_1$ components were observed in the SC state, the faster (slower) component denoted by red open (closed) squares; the red broken curve below $T_{\rm SC}$ represents the temperature dependence calculated assuming a line-node gap with $\Delta_0 / k_{\rm B}T_{\rm SC} = 2.3$.  The SC anomaly observed from the FM signal is clear evidence that superconductivity occurs in the FM region.
(c) Pressure-temperature phase diagram of UCoGe \cite{pap:Slooten2009}. Ferromagnetism is denoted as blue area, superconductivity is as yellow area. $T_{\rm SC}$($p$) extrapolates to a FM quantum critical point at the critical pressure $p_{\rm c}$. Superconductivity coexists with ferromagnetism below $p_{\rm c}$ as blue-yellow hatched area. Inset: Amplitude of $\chi_{\rm ac}(T)$ at $T_{\rm C}$ as a function of pressure. The data follow a linear $p$ dependence and extrapolate to $p_{\rm c} = 1.46 \pm 0.10$ GPa.  }
\end{figure}


\item
The temperature variation of the $^{59}$Co-NQR spectra shown in Fig. \ref{fig2} indicates that the FM region grows continuously around $T_{\rm Curie}$ coexisting with the PM region, and that the FM moment in the FM region appears discontinuously, whereas the hysteresis behavior was not observed in the temperature variation of the spectra and $1/T_1$ in the both states below $T_{\rm Curie}$ is almost identical \cite{HattoriPhysicaC}. 
Since the divergence of $1/T_1$ and the clear anomaly in the specific heat were observed at $T_{\rm Curie}$ \cite{pap:Ohta2010,pap:Aoki_review}  and the FM ordered moment is small ($\mu_s \sim 0.05 \mu_B$) at low temperatures \cite{pap:Aoki_review} , the FM phase transition in UCoGe is considered to possess weak first-order character with critical FM fluctuations, which is close to second-order one. Although we need further investigation of sample and pressure dependence on the FM transition to understand the transition character thoroughly, the presence of the critical fluctuations implies the validity of the scenarios of spin fluctuation-induced superconductivity.  

\begin{figure}[h]
\includegraphics[width=18pc]{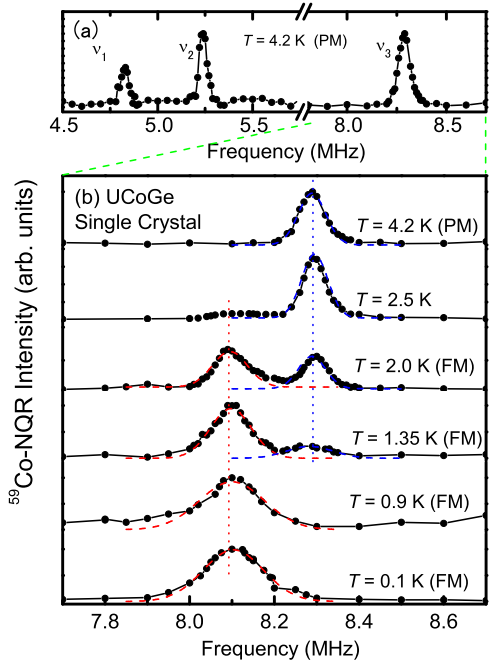}\hspace{2pc}%
\begin{minipage}[b]{16pc}\caption{\label{fig2} (a) $^{59}$Co-NQR spectra ($\nu_1, \nu_2, \nu_3$) at 4.2 K in the PM state\cite{HattoriPhysicaC}. (b) Temperature dependence of the NQR spectrum from the $\pm5/2 \leftrightarrow \pm7/2$ transitions ($\nu_3$)\cite{HattoriPhysicaC}. The PM and FM signals coexist between 1 and 2.7 K. The blue (red) broken lines represent Gaussian fits to the PM (FM) peaks; the solid lines are guides to the eye. }
\end{minipage}
\end{figure}

\item
Although transport properties show 
rather three dimensional isotropic characters~\cite{pap:Hattori2012}, studies of the SC upper 
critical magnetic field ($H_{c2}$) [Fig. 2 (a)]  and angle dependence of $H_{c2}$ [Fig. 2 (b)] 
have shown that SC properties are highly anisotropic~\cite{pap:Huy2008,
pap:Slooten2009,pap:Aoki2009}; 
the superconductivity survives in fields exceeding 15 T 
along the $a$ and $b$ axes, whereas $H_{c2}$ for fields along the $c$ direction ($H_{c2}^c$) is as small as 0.8 T.
\begin{figure}[h]
\begin{center}
\includegraphics[width=28pc]{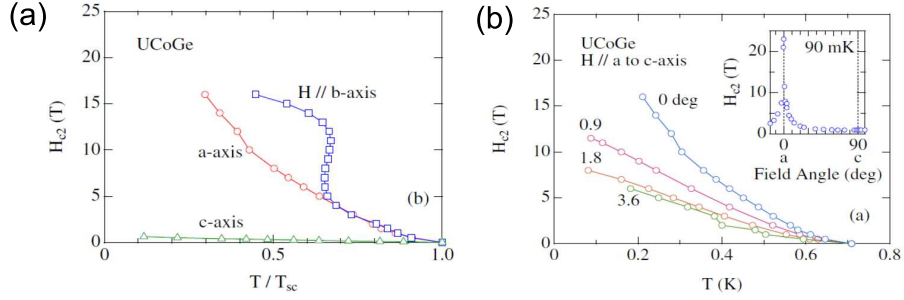}
\end{center}
\caption{\label{fig3}(a) Temperature dependence of the upper critical fields for $H \parallel a, b$ and $c$-axis~\cite{pap:Aoki2009}. The temperature is normalized by the SC critical temperature $T_{\rm SC}$ at zero field. 
(b) Temperature dependence of the upper critical field $H_{\rm c2}$ close to $H \parallel$ $a$-axis~\cite{pap:Aoki2009}. The field direction is tilted from the $a$ to $c$-axis. The inset shows the angular dependence of $H_{\rm c2}$ at 90 mK.
The values of $H_{\rm c2}$ greater than 16 T are the results of linear extrapolations to 90 mK.}
\end{figure}


\item
The static magnetic susceptibility is strongly anisotropic~\cite{pap:Huy2008}; 
the magnetic anisotropy is Ising like with the $c$ axis 
as a magnetic easy axis [Fig.~\ref{Fig4} (a)]. 
In addition, direction-dependent nuclear-spin lattice relaxation rate 
($1/T_1$) measurements on a single crystalline sample 
have revealed the magnetic fluctuations in UCoGe to be 
Ising-type FM ones along the $c$ axis (longitudinal-mode fluctuations) as shown in Fig.~\ref{Fig4} (b)~\cite{pap:Ihara2011}. 
The Ising-type fluctuations are also suggested from the anisotropic 
correlation length measured by the inelastic neutron measurements \cite{pap:Stock2011}. 
\begin{figure}[h]
\begin{center}
\includegraphics[width=28pc]{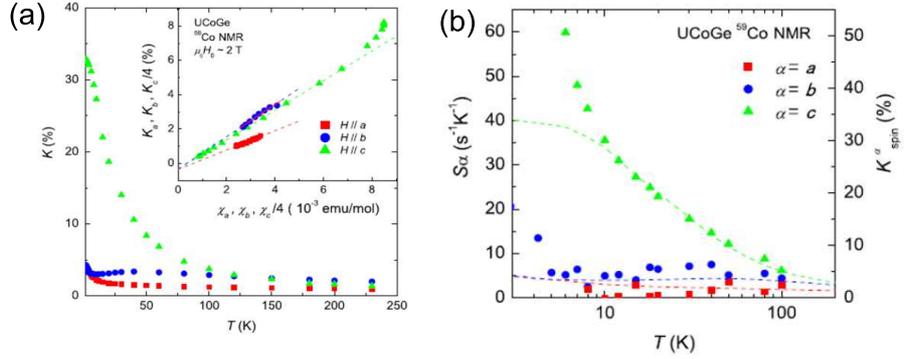}
\end{center}
\caption{\label{Fig4}NMR shift $K$ measured in field along the $a$, $b$ and $c$ directions~\cite{pap:Ihara2011}. Remarkable anisotropy was 
observed. In the inset, $K$ is plotted against the bulk susceptibility $\chi$ measured in 2 T. The nearly 
identified slopes for all the directions indicate that the hyperfine coupling constants are isotropic. (b) 
Direction-decomposed dynamic susceptibility $S_{\alpha}$ 
and static susceptibility $K_{\rm 
spin}$ along each direction~\cite{pap:Ihara2011}. Identical Ising anisotropy for both quantities found above 8 K suggests that 
the longitudinal mode dominates these fluctuations. It is noteworthy that in this $T$ range, the Knight shift scales with $S_{\alpha}$, which is predicted by the self-consistent renormalization theory for a 3-dimensional nearly FM metals~\cite{pap:Moriya1991}.    }
\end{figure}


\item
From detailed angle-resolved NMR and Meissner measurements~\cite{pap:Hattori2012}, 
it was found that magnetic fields along the $c$ axis ($H^c$) is a tuning parameter of the longitudinal FM fluctuations and strongly suppresses the FM fluctuations, as shown in Fig. 5 (a) and 5 (b). 
This sharp angle dependence of $1/T_1$ was also confirmed at 600 mK just above $T_{\rm SC}$~\cite{Hattori}. It was also shown that the superconductivity is observed in the limited 
magnetic-field region where the longitudinal FM spin fluctuations are active [Fig. 4 (c)]. 
These results suggest that the superconductivity in UCoGe is tightly
coupled with the longitudinal FM spin fluctuations along the $c$-axis.
\begin{figure}[h]
\includegraphics[width=18pc]{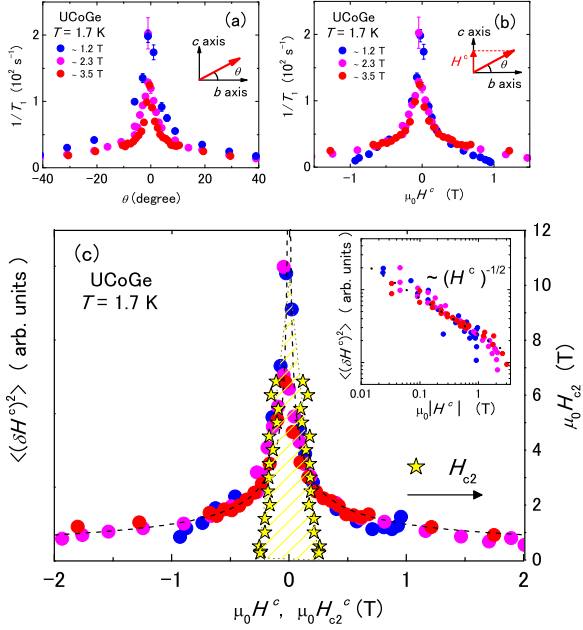}\hspace{2pc}%
\begin{minipage}[b]{16pc}\caption{\label{Fig5}(a) Angle dependence of $1/T_1$ in the $bc$
plane measured in three different magnetic fields at 1.7 K~\cite{pap:Hattori2012}.
(b) Plot of the $1/T_1$ against $H^c$. The $1/T_1$ data collapse onto a
single curve when plotted against $H^c$. 
(c) $H^c$ dependence of magnetic fluctuations along the $c$ axis $\left<(\delta H^c)^2\right>$ at 1.7 K, extracted using
$\left<(\delta H^c)^2\right> \propto \frac{1}{\cos^2{\theta}}\left(\frac{1}{T_1}(\theta)-\frac{(1+\sin^2{\theta})}{2}\frac{1}{T_1^c}\right)$.
Angle dependence of $H_{\rm c2}$, determined by the onset of the Meissner signal, is plotted against 
$H^c_{\rm c2}\equiv H_{\rm c2}\sin{\theta}$. It is found that the superconductivity is observed in the limited field region (yellow slashed area), where the longitudinal fluctuations are active. Inset: plot of $\left<(\delta H^c)^2\right>$ against $|H^c|$. The relation of $\left<(\delta Hc)^2\right> \propto 1/\sqrt{H^c}$ is shown by dashed lines in (c) and the inset. }
\end{minipage}
\end{figure}
\end{enumerate}

From a theoretical point of view, when an itinerant FM superconductor 
possesses the large energy splitting between the majority and minority spin Fermi surfaces, 
exotic spin-triplet superconductivity is anticipated, 
in which pairing is between parallel spins within each spin Fermi surface. 
In addition, it has been argued that critical FM 
fluctuations near a quantum phase transition could mediate 
spin-triplet superconductivity~\cite{pap:Monthoux1999,pap:Kirkpatrick2001,pap:Wang2001,pap:Roussev2001,
pap:Chubukov2003,pap:Fujimoto2004}, and actually, (iv) in the above indicates that
the superconductivity in UCoGe is intimately coupled with the Ising FM spin 
fluctuations.
However, as far as we know, there have been no reports 
identifying magnetic fluctuations inducing unconventional superconductivity.
Here, we show, from model calculations on the basis of the scenario
of FM spin fluctuation-mediated superconductivity in UCoGe,
that  unconventional temperature and angle dependence of $H_{c2}$ 
is consistently understood by adopting the characteristic longitudinal FM fluctuations revealed experimentally.
Besides, by comparing calculations for possible pairing states,
we also discuss candidates of SC gap symmetries in UCoGe.

\section{Calculations and Results}
\subsection{Model}
In this section, we introduce a simple model to study $H_{c2}$ in UCoGe
at ambient pressure where the ferromagnetism and superconductivity coexist.
Although UCoGe has a complicated multi-band structure,
since the spin fluctuations dominate the low energy physics in UCoGe,
we can use a simplified model to study its superconductivity
with appropriately taking into account
the critical FM spin fluctuations observed in the experiments~\cite{pap:Hattori2012,pap:Ihara2011,pap:Stock2011}.
The model relevant to UCoGe 
in the FM phase with an applied magnetic field $\vecc{H}$ is 
\begin{align}
S&=S_0+S_{\rm ex}+S_{\rm int},
\label{eq:action}\\
S_0&=\sum_{k}\int_0^{1/T}d\tau c^{\dagger}_k(\tau)[\partial_{\tau}
+\varepsilon_k]c_k(\tau),\\
S_{\rm ex}&=-(\vecc{h}_{ex}+\mu_B\vecc{H})\cdot \sum_k\int_0^{1/T}d\tau
c^{\dagger}_k(\tau)\vecc{\sigma}c_k(\tau),\\
S_{\rm int}&=-\frac{2g^2}{3}\sum_q
\int_0^{1/T}d\tau\int_0^{1/T}d\tau^{\prime} S_q^c(\tau)\chi^c(\vecc{q},
\tau-\tau^{\prime})S_{-q}^c(\tau^{\prime}),
\end{align}
and effects of the vector potential on electron motions are taken
into account in a semiclassical way.
$c_k=(c_{k\uparrow},c_{k\downarrow})^T$ is an annihilation operator
of low energy quasiparticles which are formed through hybridizations
between conduction electrons and $f$-electrons.
A typical energy scale of the quasiparticles is $40\sim 50$ (K)
estimated from the specific heat~\cite{pap:Huy2007} 
and the coherence temperature~\cite{pap:Ihara2011}.
Since the anisotropy in the resistivity in UCoGe is small as discussed in the introduction,
we use an isotropic  dispersion $\varepsilon_k=-2t\sum_{i=a,b,c}
\cos k_ia_0-\mu$ where $a_0$ is the lattice constant
and $\mu$ is the chemical potential.
We fix the electron density per site as $n=0.15$ which leads to
a small but nearly spherical Fermi surface.
As discussed in Ref.\cite{pap:Hattori2012}, the anisotropic mass model
cannot explain the anomalous angle dependence of $H_{c2}$.
In our model, the anisotropy in $H_{c2}$ arises from the exchange term and the
pairing interaction term in Eq.(\ref{eq:action}).
We use an energy unit $t=1$ and a length unit $a_0=1$.

The second term of Eq.(\ref{eq:action}) includes the exchange splitting of the Fermi surface
in the FM phase which is assumed to be large compared to the SC transition temperature at zero-field $T_{\rm SC0}$.
The exchange field is parallel to the $c$-axis and 
fixed to be $\vecc{h}_{ex}=(0,0,h_{ex})
=(0,0,0.5t)$, which would lead to suppression of the Pauli depairing 
effect against in-plane magnetic fields 
in the equal spin pairing states of the superconductivity~\cite{pap:Mineev2010}.
This is a key assumption to understand the large anisotropy in $H_{c2}$
in the present study,
because the equal spin pairing states are not robust against
the in-plane Zeeman field while the observed $H_{c2}^{a,b}$ are huge
in UCoGe.
However, we put a remark on this assumption.
UCoGe has a multi-band structure with a strong spin-orbit interaction 
and 5$f_{5/2}$ states are located near the Fermi level. 
In such a system, it is possible that, even if the magnetization in 
the FM state is small, the exchange splitting for each Fermi surface is large.
Actually, in the band calculations, the orbital magnetizations and
the spin polarizations cancel to result in a small total magnetic
moment.
The present assumption can be checked by NMR Knight shift measurements in the SC states.

Finally, the last term in Eq. (\ref{eq:action}) describes the interaction
between the quasiparticles through the Ising spin fluctuations, and
$S^c_q(\tau)=\sum_kc^{\dagger}_{k-q}(\tau)(\sigma^c/2)c_{k}(\tau)$.
The dynamical susceptibility $\chi^c(\vecc{q},i\Omega_n)$ strongly depends on the
$c$-axis magnetic field as revealed in the NMR experiments~\cite{pap:Hattori2012}.
\begin{align}
\chi^c(\vecc{q},i\Omega_n)&=\frac{\chi_0}{\delta+q^2+|\Omega_n|/\gamma_q},\\
\delta(H^c)&=h_{ex}^2+c^{\prime}\sqrt{H^c},
\end{align}
where $\gamma_q=vq$ and $v=4$ is approximately the Fermi velocity.
Here, we have assumed a simple Landau damping in $\chi^c(\vecc{q},i\Omega_n)$,
although existence of non-Landau damping was reported~\cite{pap:Stock2011,pap:Mineev2012}.
The coefficient $c^{\prime}$ is determined by the NMR experiments,
and a normalization factor $H_0^c\sim 1$ T is introduced as 
$c^{\prime}\equiv ch_{ex}^2/\sqrt{H^c_0}$ so that the dimensionless
parameter $c$ is $c\sim {\mathcal O}(1)$.
Although the origin of this magnetic field dependence 
in the susceptibility is not so clear, 
we have a possible explanation for it. 
According to the band calculation, the 5$f_{5/2}$ bands in UCoGe 
have some semi-metallic like structures around several 
$k$-points~\cite{pap:Samsel2010}. 
For such structures, the density of states would be $\sim \sqrt{H^c}$, 
which results in $(\chi^c)^{-1}\sim {\rm const.}+\sqrt{H^c}$
near the criticality. 
In this scenario, detailed structures in the bands are important, 
and therefore, the spin fluctuations in UCoGe under a magnetic field 
can depend on sample qualities.
This problem will be studied elsewhere.
We note that the coupling constant $g$ should be regarded as 
a renormalized one including the vertex corrections which
enhance the SC transition temperatures.
Although we have postulated a simple analytic $q$-dependence in
$\chi^c$, there might exist non-analytic corrections
$\sim q^2\log q$ in 3-dimensional isotropic systems
\cite{pap:Belitz,pap:Chubukov,pap:Duine2005,pap:Conduit2009}.
The non-analytic corrections are important in connection with fluctuation-induced first order phase transitions. 
However, 
since the first order character of the FM transition is weak, and critical FM fluctuations are well developed 
in UCoGe as discussed in Sec.\ref{sec:intro},
it would be legitimate to neglect the first order character, and assume that 
$\chi^c$ is analytic in $q$ as long as calculations of the superconducting properties are concerned.
In addition, if exist in the Ising-like system, the non-analytic corrections would not change the present
study qualitatively as discussed in~\cite{pap:Tada2011}.

\subsection{Eliashberg equation}
In this study, we focus on the anisotropy between the $a$ and $c$-axes
upper critical fields ($H_{c2}^a$ and $H_{c2}^c$). However,
it was reported that the $b$-axis upper critical field $H_{c2}^b$ shows a "S-shaped" curve
at high fields, and relations to the reentrant superconductivity
in URhGe were discussed~\cite{pap:Aoki2009}.
In URhGe, the magnetization flips at a critical $H^b=H^{b\ast}$ and the
spin fluctuations at $H^{b\ast}$ are considered to be different from
those of $H^b=0$~\cite{pap:Levy2005}.
Similar scenarios are expected to hold also in UCoGe where
the S-shaped $H_{c2}^b$ would indicate a crossover of the superconductivities
of different mechanisms under magnetic field along the $b$ axis.
The problem of $H_{c2}^b$ in UCoGe is an another issue and will be
studied elsewhere.

To determine the upper critical field $H_{c2}$ in the $ac$-plane, 
we solve the linearized Eliashberg equation in the presence of 
a vector potential~\cite{pap:Tada2010}. 
The Eliashberg equation for an equal-spin pairing state in the FM phase is
\begin{align}
\Delta_{\sigma \sigma}(k)=&-\frac{T}{2N}
\sum_{k^\prime}V(k,k^{\prime})[
G_{\sigma\sigma^{\prime}}(k^{\prime}+\Pi)G_{\sigma\sigma^{\prime}}(-k^{\prime})
+G_{\sigma\sigma^{\prime}}(k^{\prime})G_{\sigma\sigma^{\prime}}(-k^{\prime}+\Pi)]
\Delta_{\sigma^{\prime}\sigma^{\prime}}(k^{\prime}),
\end{align}
where $\Pi=(-i\nabla_R-2e\vecc{A}(\vecc{R}),0)$ and 
$\nabla\times\vecc{A}=\vecc{H}=H(\cos\theta,0,\sin\theta)$.
A short notation $k=(\vecc{k},i\omega_n)$ is used hereafter.
Here we have neglected the effect of the magnetization in 
the orbital motion, because the observed moment is very small in 
UCoGe~\cite{pap:Mineev2002}.
Note that the Green's function $\hat{G}(k)$ has off-diagonal elements
when $\theta\neq 90^{\circ}$. The gap function depends both on the momentum
in the relative coordinate and the center-of-mass coordinate, and
is expanded as $\Delta_{\sigma\sigma}(k,\vecc{R})=
\Delta_{\sigma\sigma}(k)\cdot \phi_0(\vecc{R})$, where $\phi_0$
is the lowest level Landau function.

We solve the linearized Eliashberg equation for two cases corresponding
to the classification by the magnetic point group $D_2(C_2^z)$
for UCoGe which has an orthorhombic crystal structure~\cite{pap:Mineev2002,
pap:Mineev_sym}.
One is the so-called A state which corresponds to 
point node symmetry with a $d$-vector $\vecc{d}\sim (a_1k_a+ia_2k_b,a_3k_b+
ia_4k_a,0)$ near the $\Gamma$ point. The other one is the so-called B
state corresponding to horizontal line node symmetry with
$\vecc{d}\sim (b_1k_c+ib_2k_ak_bk_c,ib_3k_c+b_4k_ak_bk_c,0)$.
Here, $\{a_i\}$ and $\{b_i\}$ are real coefficients.
Both of the gap functions are determined by self consistent calculations,
and the resulting SC states are non-unitary
because of the presence of the exchange field $\vecc{h}_{ex}$.
We note that, under the applied magnetic field $\vecc{H}$,
the calculated gap functions are distorted and has lower symmetry
than that of the cubic lattice in the present model.

The pairing interaction is evaluated at the first order in 
$g^2\chi_0$,
\begin{align}
V(k,k^{\prime})=-\frac{g^2}{6}\chi^c(k-k^{\prime})+\frac{g^2}{6}
\chi^c(k+k^{\prime}).
\end{align}
For the selfenergy, we include only the term
\begin{align}
\Sigma_0(k)&\equiv (\Sigma_{\uparrow\uparrow}(k)
+\Sigma_{\downarrow\downarrow}(k))/2\notag \\
&=\frac{T}{6N}\sum_qg^2\chi^c(q)[G^0_{\uparrow\uparrow}(k+q)+
G^0_{\downarrow\downarrow}(k+q)],
\end{align}
where $G^0_{\sigma\sigma}$ is the non-interacting Green's function.
Other neglected terms in the self energy are much smaller than
$\Sigma_0$ in magnitude.
As noted in the previous section, the coupling constant $g$ is 
a renormalized one.

In the present study, the coupling constant is fixed as $g^2\chi_0=125t$
which is rather large to give a high transition temperature so that
the numerical costs are reduced.
However, qualitatively, our main results do not depend on the value of
$g^2\chi_0$. 
For this value of $g^2\chi_0$, the mass enhancement factor $1/z$ is $\sim 1.3$,
which is not so large because the density of states at the Fermi surface is small
in the present parameters.
The transition temperature for the A state without magnetic
field is $T_{\rm SC0}=0.020t$ which we regard as 1.0 K according to the
experiments for UCoGe. Then, the hopping integral is fixed as 
$t=50$ K throughout in our study, and the order of this value is
consistent with the coefficient of the linear specific heat
$\gamma\sim 50$ mJ/K$^2$mol~\cite{pap:Huy2007} and the coherence temperature 
$T^{\ast}\sim 40$ K~\cite{pap:Ihara2011}. We also fix the lattice constant,
$a_0=4.0$ \AA which leads to an effective mass for the orbital motion
of the Cooper pairs $m_{\rm eff}\equiv \hbar/ta_0^2\sim 440m_0$ where
$m_0$ is the bare electron mass.
Correspondingly, we fix $H_0^c=0.02t\sim 1$ T, which is consistent
with the NMR experiments~\cite{pap:Hattori2012}.

\subsection{Results for A state}
Before discussing numerical results of $H_{c2}$ of the present model,
we note that the orbital limit $H_{\rm orb}$ depends
on positions of gap nodes in general~\cite{pap:Tada2011}. A simple explanation is as follows.
Let us consider a superconductor with isotropic Fermi
surface which has nodes in a gap function.
The effective
velocity for the cyclotron motion of the orbital depairing
effect can be of the form of $\tilde{\vecc{v}}_k = \varphi_k\vecc{v}_k$ 
for the basis function 
$\varphi_k$ corresponding to the SC gap symmetry,
where $\vecc{v}_k$ is a velocity of electrons.
For $\varphi_k$ with horizontal line nodes,
when the magnetic field is parallel to the $c$-axis,
the cyclotron
motion is suppressed at the nodes where in-plane $\tilde{\vecc{v}}_k$
is zero,
resulting in a large orbital limiting field.
The cyclotron motion is not suppressed for $H\perp c$
because $\tilde{\vecc{v}}_k\perp H$ is non-zero on the Fermi
surface, which leads
to the anisotropy of the orbital limiting field $H_{\rm orb}^{\parallel c}>
H_{\rm orb}^{\perp c}$. On
the other hand, for the point node gap function,
$\tilde{\vecc{v}}_k$ cannot be zero
except for the poles of the Fermi surface
$k_a = k_b = 0$ which are realized as cross points of two line nodes
$k_a=0$ and $k_b=0$. Therefore, for point nodes at the poles of the
Fermi surface, the order is turned over, $H_{\rm orb}^{\parallel c}<
H_{\rm orb}^{\perp c}$.
Following this consideration, if there is no suppression of the
pairing interaction and the Pauli depairing effect is negligible, 
$H_{c2}^c<(>)H_{c2}^a$ is expected for the A (B) state.

\begin{figure}[h]
\begin{minipage}[t]{17pc}
\includegraphics[width=17pc]{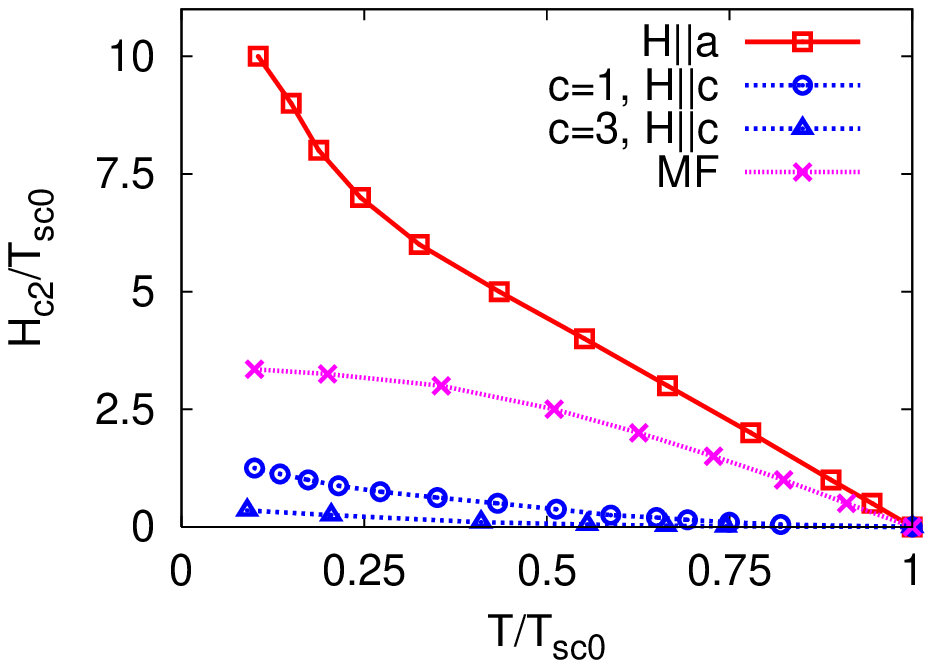}
\caption{\label{fig:A1}Temperature dependence of $H_{c2}$ for
the $a$ and $c$ axes in the A state. 
The purple curve "MF" is $H_{c2}^c$ calculated with 
the mean field $\delta(H^c)$.}
\end{minipage}\hspace{2pc}%
\begin{minipage}[t]{17pc}
\includegraphics[width=17pc]{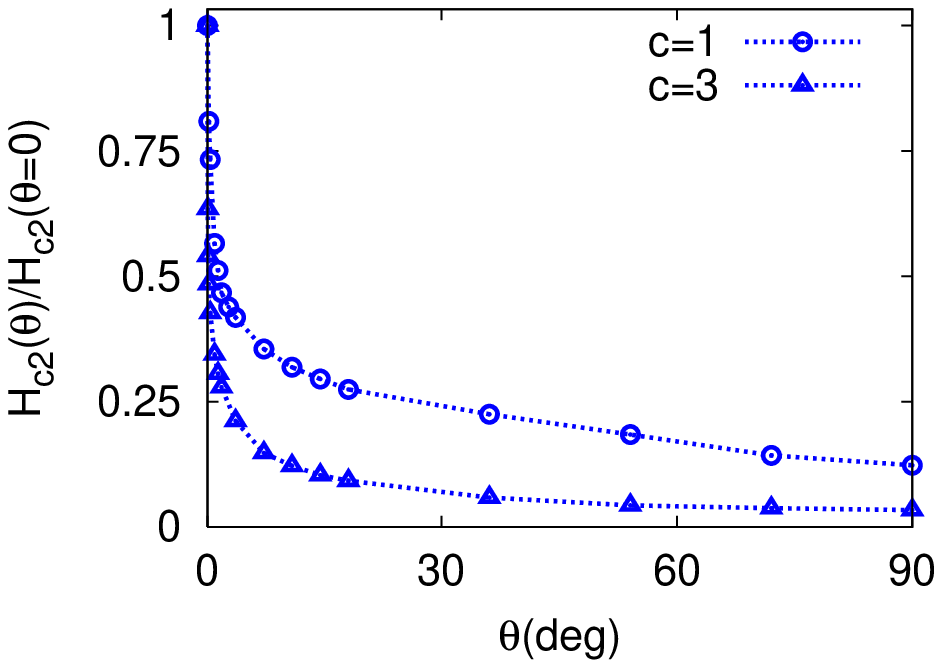}
\caption{\label{fig:A2}Field angle dependence of $H_{c2}$ at 
$T=0.1T_{\rm SC0}$ in the A state.}
\end{minipage} 
\end{figure}

We now turn to discussions of calculation results.
Temperature dependence of $H_{c2}^{a,c}$ with $c=1,3$ 
for the A state is shown in Fig. \ref{fig:A1}.
It is found that the qualitative behaviors of $H_{c2}$ are independent of 
$c$. 
$H_{c2}^a$ exhibits a strong coupling behavior and reaches $H_{c2}^a/T_{\rm SC0}\sim 10$.
Although the precise value of $H_{c2}^a/T_{\rm SC0}$ depends on the choice of $h_{ex}$ and
$g^2\chi_0$, strong coupling calculations of $H_{c2}$ naturally explain the upward curvature
in $H_{c2}^a$ and its large value observed in the experiments, which is a key to understand
the anomalous anisotropic behavior in $H_{c2}$.
The calculated anisotropy ratio $H_{c2}^a/H_{c2}^c$ becomes $10\sim20$ 
in consistent with the experiments, although
the Fermi surface of our model is merely spherical.
We emphasize that, for $H_{c2}$ to have a large anisotropy between 
the $a$ and $c$-axes, the experimentally observed $\sqrt{H^c}$-dependence 
in the susceptibility is crucially important. 
For a comparison, we calculate $H_{c2}^c$ in the case that
$\chi^c$ has weaker $H$-dependence, mean-field like dependence of
$\delta(H^c)=(h_{ex}+\mu_BH^c)^2$, 
for which results are shown with purple symbols in Fig. \ref{fig:A1}.
In this case, suppression of $H_{c2}$ for the $c$-axis is only moderate, 
and we cannot obtain anomalously large anisotropy in $H_{c2}$. 
In Fig. \ref{fig:A2}, we show field angle dependence of $H_{c2}(\theta)$
at $T=0.1T_{\rm SC0}$, where $\theta$ is the angle from the $a$-axis to
the $c$-axis.
It is seen that $H_{c2}(\theta)$ is strongly suppressed when
the field direction is slightly tilted from the $a$-axis to the
$c$ axis.
This strong angle dependence is 
again due to the $\sqrt{H^c}$ suppression of the pairing
interaction.
The calculation results well explain the unusual angle dependence of $H_{c2}$ observed experimentally.
This good agreement is strong evidence that the superconductivity
in UCoGe is actually mediated by the critical Ising spin fluctuations.

We note that
the relation between $H^c$ and $H_{c2}$ is considered to be an analog of the isotope effect
in conventional superconductors where ion mass $M$ controls 
transition temperature $T_{\rm SC}$. 
In both cases, suppression of the pairing interaction by increasing $H^c$ or $M$
leads to reduction of $T_{\rm SC}$. 
One important reason why we can establish such a relation between $H^c$
and $H_{c2}$ in 
the $f$-electron system UCoGe is that the critical spin fluctuations
dominate the low energy physics.
Details of the system and complexity of multi-band
are sub-dominant factors for the superconductivity, and therefore,
the SC properties are well understood once the spin fluctuations
are appropriately taken into account.

\subsection{Results for B state}
\begin{figure}[h]
\begin{minipage}[t]{17pc}
\includegraphics[width=17pc]{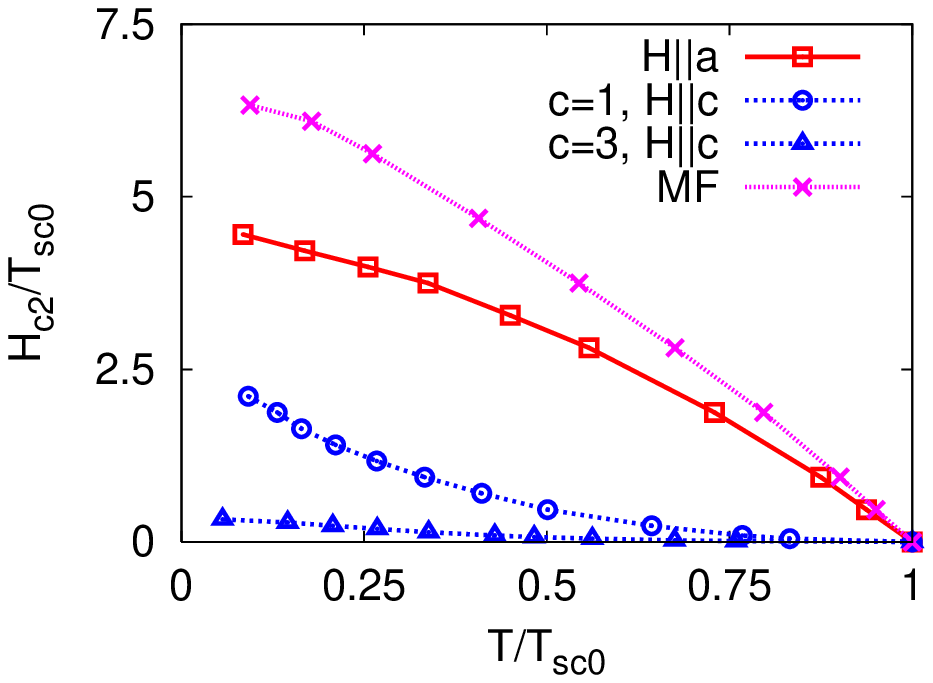}
\caption{\label{fig:B1}Temperature dependence of $H_{c2}$ for
the $a$ and $c$ axes in the B state.
The purple curve "MF" is $H_{c2}^c$ calculated with 
the mean field $\delta(H^c)$. }
\end{minipage}\hspace{2pc}%
\begin{minipage}[t]{17pc}
\includegraphics[width=17pc]{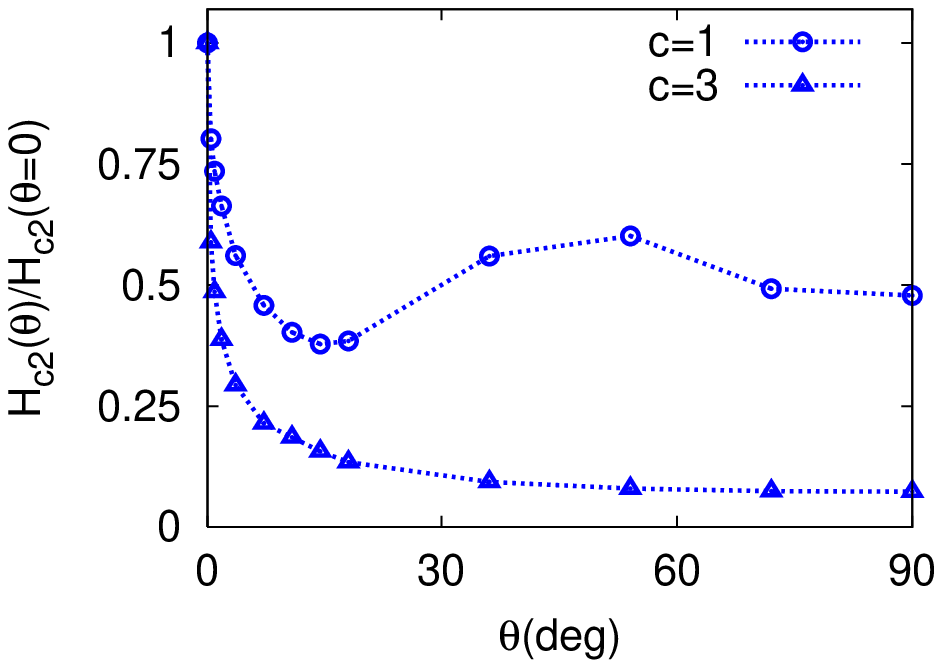}
\caption{\label{fig:B2}Field angle dependence of $H_{c2}$ at 
$T = 0.1 T_{\rm SC0}$ in the B state.}
\end{minipage} 
\end{figure}

The transition temperature at zero-field for the B state is
$T_{\rm SC0}=0.0213t$ which differs from that for the A state, 
because of the presence of the exchange field along the $c$-axis. 
However, the difference is not important in our study. 
Temperature dependence of $H_{c2}^{a,c}$ for the B state is shown in Fig. \ref{fig:B1}.
For the B state, strong-coupling behaviors for the $a$-axis are not so 
significant, because the orbital limiting fields depend on 
the positions of gap nodes as discussed in the previous section,
and the enhancement of 
$H_{c2}^a$ for the B state is weaker than that for the A state. 
For the present value of $m_{\rm eff}$, the Pauli depairing effect 
cannot be negligibly small at high magnetic fields, 
which leads to the suppression of $H_{c2}^a$ in that region. 
For the $c$-axis, 
detailed behaviors depend on the value of the parameter $c$. 
In the case of rather weak suppression of the Ising fluctuations, 
i.e. $c = 1$, $H_{c2}$ is enhanced at low temperatures. 
If the suppression is sufficiently strong, 
for example $c = 3$, such enhancement cannot be seen.
We also show $H_{c2}^c$ calculated with the mean field $\delta(H^c)$
by the purple curve as in the A state.
In the B state, $H_{c2}^c$ is not strongly suppressed and
$H_{c2}^c$ is larger than $H_{c2}^a$ in contrast to the experiments.
The relation $H_{c2}^c>H_{c2}^a$ with the mean-field $\delta(H^c)$
is directly understood from the general relation
$H_{\rm orb}^c>H_{\rm orb}^a$ for horizontal line node gap functions
in simple systems.

Behaviors of $H_{c2}$ as a function of the field angle also depend 
on the value of $c$ as shown in Fig. \ref{fig:B2}. 
For $c = 1$, $H_{c2}(\theta)$ shows a 
non-monotonic behavior. This is because, 
for the B state, the relation $H_{c2}(\theta=0^{\circ})
<H_{c2}(\theta=90^{\circ})$ 
holds in the orbital limit if there is no suppression of the Ising 
fluctuations, while in fact the suppression in $\chi^c(q)$ gets 
stronger as $\theta\rightarrow 90^{\circ}$. 
On the other hand, for $c = 3$, $H_{c2}$ monotonically decreases. 
Comparing these results with those  for the A state which are robust 
against the parameter value, one may conclude that the A state is a 
promising candidate for the SC realized in the FM state in UCoGe, 
although the B state is not ruled out.

\section{Summary}
In this paper, we have studied the upper critical field $H_{c2}$ for the 
FM superconductor UCoGe in the FM phase.
In the introduction, we discussed several experimental observations
which are key clues to understand relations between the FM and SC.
To clarify the relationship,
we introduced the simple model which includes the critical FM spin
fluctuations with anomalous $\sqrt{H^c}$ dependence revealed by the
NMR experiments.
The linearized Eliashberg equations are solved within the first order in the
spin fluctuations.
In the A state with point nodes, temperature dependence of
$H_{c2}^a$ shows an upward curvature while $H_{c2}^c$ is suppressed.
When the field angle is tilted from the $a$ to the $c$ axis,
$H_{c2}$ is strongly suppressed.
These behaviors are totally due to the characteristic suppression
of the FM spin fluctuations by $H^c$, and in nice agreement with
the experimentally observed $H_{c2}$ in UCoGe.
This is strong evidence of the scenario that the spin-triplet
superconductivity is actually mediated by the critical Ising FM
spin fluctuations in UCoGe.
The relation between $H^c$ and $H_{c2}$ is an analog of the
isotope effect in conventional superconductors.
In the B state with a horizontal line node,
both enhancement of $H_{c2}^a$ and suppression of $H_{c2}^c$ are
moderate compared to those in the A state.
Therefore, the A state is a promising candidate for the 
pairing symmetry in UCoGe, although the B state is not ruled out.

Finally, we put remarks on some remaining problems
related to the present study.
Firstly, 
in the present study, we have simply adopted the experimentally
observed $\sqrt{H^c}$ dependence in $\chi^c(q)$.
Although we suggested a possible explanation, 
the origin of this anomalous dependence is not clear.

Secondly,
we have assumed that the exchange splitting
is large enough to suppress the Pauli depairing effect.
However, this assumption should be checked experimentally by
e.g. NMR Knight shift measurements.
There is also a related problem.
We have focused on the coexistence region of the superconductivity
and the ferromagnetism where the exchange splitting can be large.
Under pressure~\cite{pap:Slooten2009,pap:Aoki2009},
$T_{\rm Curie}$ is suppressed and the exchange splitting becomes small.
In this case and in the paramagnetic phase, 
the Pauli depairing effect would not be suppressed.
Actually, it was reported that, around the critical pressure where 
$T_{\rm Curie}$ is extrapolated to zero,
$H_{c2}^{a,b}(T\rightarrow0)$ is suppressed while 
$H_{c2}^c(T\rightarrow0)$ is increasing~\cite{pap:Aoki2012}.
These behaviors could be understood as a result of two competing
effects; although the Pauli depairing effect would not be
suppressed by the exchange field near the criticality,
the FM spin fluctuations are enhanced.

Investigations of these issues are needed for a comprehensive 
understanding of UCoGe, and they would be studied in the future.

\ack
The authors thank S. Yonezawa, and Y. Maeno for experimental support and valuable discussions, and D. Aoki, J. Flouquet, A. de Visser, A. Huxley, H. Harima, and H. Ikeda for valuable discussions. 
This work was partially supported by Kyoto Univ. LTM centre, Yukawa Institute, the "Heavy Electrons" Grant-in-Aid for Scientific Research on Innovative Areas  (No. 20102006, No. 21102510,  No. 20102008, No. 23102714, and No. 23840009) from The Ministry of Education, Culture, Sports, Science, and Technology (MEXT) of Japan, a Grant-in-Aid for the Global COE Program ``The Next Generation of Physics, Spun from Universality and Emergence'' from MEXT of Japan, a grant-in-aid for Scientific Research from Japan Society for Promotion of Science (JSPS), KAKENHI (S) (No. 20224015), KAKENHI (C) (No. 23540406) and FIRST program from JSPS .


\end{document}